\documentclass[aps,prb,superscriptaddress,reprint,notitlepage]{revtex4-2}
\usepackage[utf8]{inputenc}
\usepackage[hidelinks]{hyperref}
\usepackage{hyperref}
\usepackage{color}
\hypersetup{colorlinks=true}
\usepackage{graphicx}
\usepackage{bm}
\usepackage{amsmath}
\usepackage{amssymb}
\usepackage{xspace}
\usepackage[capitalise]{cleveref}
\usepackage{txfonts}
\graphicspath{{./figures/}}
\usepackage{physics}
\usepackage{siunitx}
\usepackage{booktabs}
\usepackage{chemformula}
\usepackage{orcidlink}

\usepackage{float}
\usepackage{comment}

\newcommand{\subfigref}[2]{Fig.~\hyperref[#1]{\ref*{#1}#2}}

\newcommand{\dg}{\dagger}
\newcommand{\up}{\uparrow}
\newcommand{\dn}{\downarrow}
\newcommand{\down}{\downarrow}

\begin{document}
\title{Triplon dynamics from partial Kondo screening in frustrated heavy-fermion systems}
\author{Hiroki Nakahara\orcidlink{https://orcid.org/0009-0006-4273-8304}}
\email{nakahara-h@fuji.waseda.jp}
\affiliation{Department of Applied Physics, Waseda University, Okubo, Shinjuku-ku, Tokyo 169-8555, Japan}

\author{Masahito Mochizuki\orcidlink{https://orcid.org/0000-0002-0128-0580}}
\email{masa\_mochizuki@waseda.jp}
\affiliation{Department of Applied Physics, Waseda University, Okubo, Shinjuku-ku, Tokyo 169-8555, Japan}

\author{Rintaro Eto\orcidlink{https://orcid.org/0000-0001-8833-1311}}
\email{rintaro.eto@tum.de}
\affiliation{Technical University of Munich, TUM School of Natural Sciences, Physics Department, 85748 Garching, Germany}
\affiliation{Department of Applied Physics, Waseda University, Okubo, Shinjuku-ku, Tokyo 169-8555, Japan}

\date{\today}
\begin{abstract}
    The partially disordered (PD) phase, in which magnetically ordered and Kondo-screened sites coexist, has been proposed as a possible ground state of geometrically frustrated heavy-fermion systems.
    We investigate its dynamical properties in the half-filled periodic Anderson model on a triangular lattice by calculating the dynamical spin susceptibility within the random-phase approximation built on a self-consistent Hartree-Fock solution. 
    We show that successive quantum phase transitions from the Kondo-insulating phase to the PD phase and to a magnetically ordered phase are driven by the condensation of triplon modes originating predominantly from the Kondo-screened sites.
    In the PD phase, magnons and triplons coexist and hybridize, leading to characteristic level repulsion in the excitation spectrum.
    These results reveal how triplet excitations evolve into collective magnetic modes across the successive phase transitions and highlight the dynamical interplay between magnetic order and Kondo screening in frustrated heavy-fermion systems.
\end{abstract}

\maketitle

\section{Introduction}
Rare-earth compounds exhibit a rich variety of correlated-electron phenomena arising from the interplay between conduction electrons and localized magnetic moments. Two competing mechanisms emerge from this interplay. One is the Kondo effect, in which conduction electrons screen localized magnetic moments through the formation of Kondo singlets~\cite{Kondo1964,Yoshida1966}. The other is the Ruderman-Kittel-Kasuya-Yosida (RKKY) interaction, whereby conduction electrons mediate effective exchange interactions between localized moments and promote magnetic ordering~\cite{RK1954,K1956,Y1957}. The competition between Kondo screening and the RKKY interaction lies at the heart of heavy-fermion physics~\cite{Doniach1977} and has long served as a paradigmatic framework for understanding emergent quantum phases and quantum criticality, including unconventional superconductivity and non-Fermi-liquid behavior~\cite{Mathur1998,Gegenwart2008}.

\begin{figure}[hbt]
    \centering
    \includegraphics[width=0.45\textwidth]{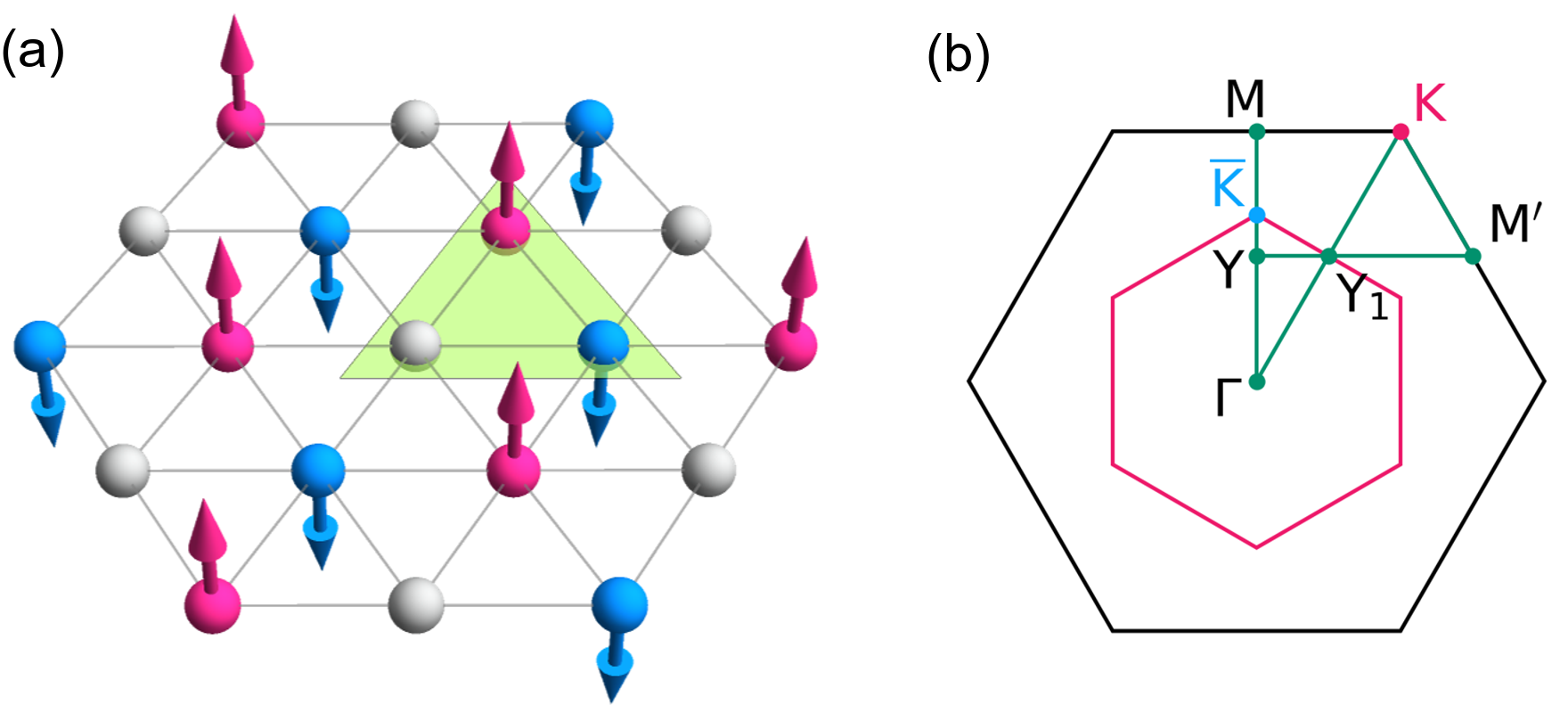}   
    \caption{
        (a) Schematic of the three-sublattice partially disordered state on the two-dimensional triangular lattice. A magnetic unit cell highlighted in green includes a Kondo-screened site without magnetic moment (gray ball) and the remaining two sites with collinear antiferromagnetic order (magenta and cyan balls with arrows indicating spin orientation).
        (b) The first Brillouin zone (black) and the magnetic Brillouin zone (magenta). Green lines with markers indicate a high-symmetry momentum path employed in this work.
    }
    \label{Fig01}
\end{figure}

Geometrical frustration can further enrich the competition between Kondo screening and magnetic ordering. While frustration in localized-spin systems is often relieved through the formation of noncollinear magnetic orders~\cite{Akagi2010} or exotic quantum states such as valence-bond solids~\cite{Affleck1987}, resonating-valence-bond states~\cite{Anderson1973}, and quantum spin liquids~\cite{Yan2011,Slagle2014}, heavy-fermion systems, in contrast, possess an additional mechanism for relieving frustration through on-site Kondo screening. This extra degree of freedom may stabilize a partially disordered (PD) state schematically shown in Fig.~\ref{Fig01}(a), in which a subset of local moments is quenched by Kondo screening while the remaining moments participate in magnetic order~\cite{Lacroix1996}. Indeed, such states have been experimentally observed in geometrically frustrated heavy-fermion compounds, including CePdAl, which hosts a distorted kagome lattice of Ce ions~\cite{Donni1996, Dolore1997, Oyamada2008, Mochidzuki2017, Fritsch2017}, and UNi$_4$B, which contains a triangular-lattice arrangement of U ions~\cite{Mentink1994, Oyamada2007, Haga2008, Hayami2015, Saito2018, Oyamada2018, Yanagisawa2021, Tabata2021, Willwater2021, Ishitobi2023, Hidaka2025, Riedel2026}. However, the microscopic origin and stability of the PD state remain subjects of active debate.

To address this issue, extensive theoretical studies have been carried out using microscopic models for heavy-fermion systems. Among them, the periodic Anderson model (PAM)~\cite{Anderson1961}, which incorporates both the hybridization between conduction and localized electrons and the on-site Coulomb repulsion of localized electrons, has served as a minimal framework for describing the competition between Kondo screening and magnetic ordering. In the limit of strongly localized $f$ electrons, the PAM reduces to the Kondo-lattice model (KLM)~\cite{SW1966}, which has also been widely employed to investigate the PD state. Various theoretical studies have reported the stabilization of the PD state as a ground state in frustrated heavy-fermion models~\cite{Motome2010, Hayami2011, Hayami2012, Noda2014, Aulbach2015, Kebler2020}. A variational Monte Carlo study has provided evidence for a PD ground state in the KLM~\cite{Motome2010}, whereas the PD state has been obtained within self-consistent Hartree-Fock calculations and cluster dynamical mean-field theory in the PAM~\cite{Aulbach2015}.

Nevertheless, a unified understanding of the stability of the PD state has yet to be established. Since the PD state is characterized by the coexistence of magnetically ordered and Kondo-screened sites, its theoretical description requires both an appropriate treatment of quantum fluctuations associated with Kondo screening and sufficient spatial resolution to capture the underlying site-dependent order. Simultaneously satisfying these requirements remains a challenging task, and consequently the stability of the PD state can depend sensitively on the theoretical approach employed.
Furthermore, understanding the dynamical properties of the PD state is equally important. While previous studies have primarily focused on its ground-state properties, collective excitations provide complementary insights into the nature and stability of the underlying order. In particular, the softening or condensation of collective modes may signal instabilities toward competing phases~\cite{josef2023,Rao2025}.

In this work, we investigate the stability and collective excitations of the PD state in the half-filled PAM on the two-dimensional triangular lattice by applying the random-phase approximation (RPA) on top of the self-consistent Hartree-Fock mean-field solution. Although quantum fluctuations are incorporated only at a perturbative level within the present framework, the RPA retains full momentum resolution and is therefore well suited for identifying spatially modulated instabilities. Moreover, dynamical response functions can be computed straightforwardly within the same framework, enabling a unified investigation of both the stability and collective excitations of the PD state.

Using this approach, we find that the transition from the Kondo-insulator phase to the PD phase is driven by the condensation of a spin-flip excitation of the Kondo singlet~\cite{G.M.Zhang2000,Christoph2001}, namely a triplon. Specifically, two of the three sublattices develop collinear N\'{e}el order, whereas the remaining sublattice remains Kondo-screened. We further elucidate the evolution of the collective excitation spectrum in the PD phase and identify a previously unexplored instability toward a nine-sublattice magnetic order. Furthermore, we find hybridization and level repulsion between triplon and magnon modes in the PD phase, revealing a nontrivial coupling between magnetically ordered and Kondo-screened sites. Our results, obtained within a perturbative framework, demonstrate that collective low-energy excitations play an essential role in stabilizing the PD state and provide characteristic dynamical signatures of the coexistence of magnetic order and Kondo singlets.

\section{Model and Methods}
\begin{figure*}[hbt]
    \centering
    \includegraphics[width=1\textwidth]{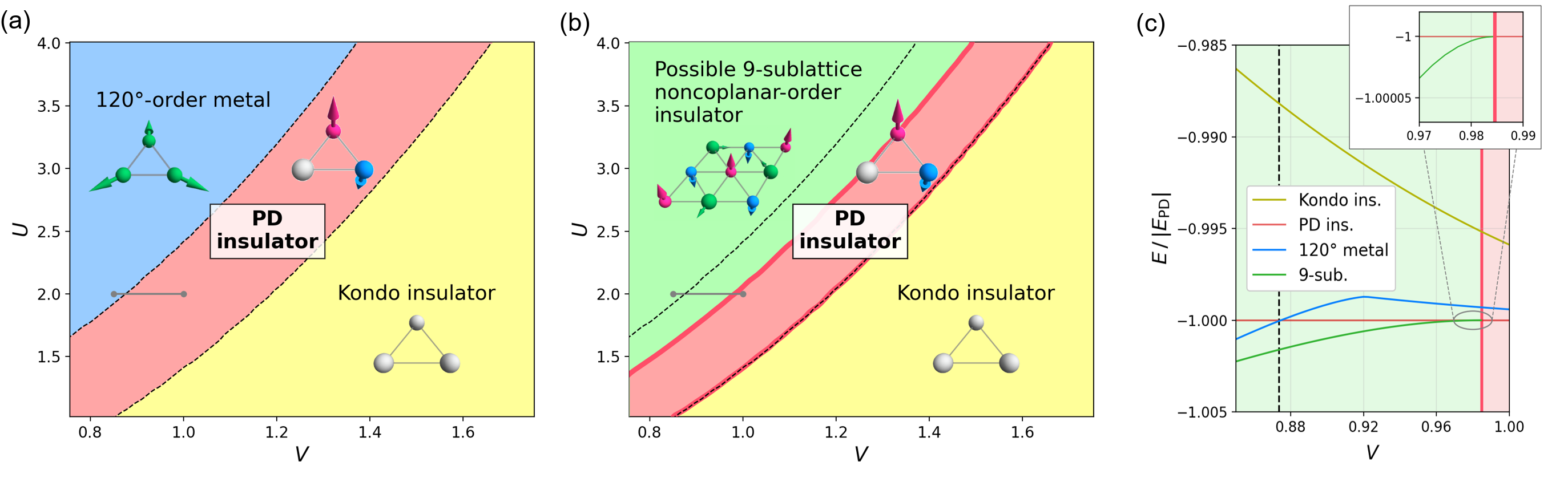}   
    \caption{
        (a) SCHF phase diagram obtained by restricting the magnetic unit cell to three sublattices, as reported in Ref.~\cite{Hayami2011}. 
        (b) SCHF+RPA phase diagram obtained by identifying linear instabilities of the SCHF mean-field states within RPA. While the KI-PD phase boundary coincides with that obtained by SCHF, SCHF+RPA reveals an additional instability of the three-sublattice PD state toward a larger-unit-cell magnetic state, which is absent in the SCHF phase diagram restricted to the three-sublattice unit cell. 
        (c) Comparison of the free energies of different magnetic states as a function of the hybridization $V$ along the $U=2$ cut indicated by the horizontal gray line in (a) and (b). The vertical axis shows the ratio of the free energy of each state to that of the PD phase at each $V$. Among these states, the state with the lowest free energy is regarded as the ground state within SCHF, and the corresponding phase is indicated by the background color. The inset shows a magnified view near the transition.
        In (a) and (b), the black dashed lines indicate the phase boundaries determined by the SCHF free energy, whereas the thick red lines indicate those identified by collective mode condensations within RPA. Along the $U=2$ cut, the phase boundary between the PD and the nine-sublattice order obtained by SCHF+RPA agrees with the crossing of the corresponding SCHF free energies shown in (c).
    }
    \label{Fig02}
\end{figure*}
\subsection{Periodic Anderson Model}
We consider the periodic Anderson model on a two-dimensional triangular lattice, whose Hamiltonian is given by
\begin{align}
\label{eq:PAM_Hamiltonian}
\hat{\mathcal{H}}_{\mathrm{PAM}}
=
&-t\sum_{\langle\vb{r},\vb{r}'\rangle,\sigma}
\left(
\hat{c}^{\dagger}_{\vb{r}\sigma}
\hat{c}_{\vb{r}'\sigma}
+\mathrm{h.c.}
\right)
\nonumber\\
&
-V\sum_{\vb{r},\sigma}
\left(
\hat{c}^{\dagger}_{\vb{r}\sigma}
\hat{f}_{\vb{r}\sigma}
+\mathrm{h.c.}
\right)
\nonumber\\
&
+E_f\sum_{\vb{r},\sigma}
\hat{f}^{\dagger}_{\vb{r}\sigma}
\hat{f}_{\vb{r}\sigma}
+
U\sum_{\vb{r}}
\hat{f}^{\dagger}_{\vb{r}\uparrow}
\hat{f}_{\vb{r}\uparrow}
\hat{f}^{\dagger}_{\vb{r}\downarrow}
\hat{f}_{\vb{r}\downarrow}.
\end{align}
Here, $\hat{c}^{\dagger}_{\vb{r}\sigma}$ and $\hat{f}^{\dagger}_{\vb{r}\sigma}$ are the creation operators for conduction and localized $f$ electrons with spin $\sigma~(=\uparrow,\downarrow)$ at lattice site $\vb{r}$, respectively. The first term describes the kinetic energy of the conduction electrons, where only nearest-neighbor hopping is taken into account. Throughout this work, we take the nearest-neighbor hopping amplitude $t$ as the unit of energy. The second term represents the onsite hybridization between the conduction and localized electrons, which favors the formation of Kondo singlets. The third term represents the localized $f$-electron energy level. The last term describes the onsite Coulomb repulsion between localized $f$ electrons.

\subsection{Hartree-Fock Approximation}
To investigate the magnetic ground state of the periodic Anderson model, we employ the self-consistent Hartree-Fock approximation. The onsite Coulomb interaction is decoupled as
\begin{equation}
\begin{aligned}
    \hat{f}^{\dg}_{\vb{r}\up}\hat{f}_{\vb{r}\up}\hat{f}^{\dg}_{\vb{r}\down}\hat{f}_{\vb{r}\down}
    &\approx
      \langle \hat{f}^{\dg}_{\vb{r}\up}\hat{f}_{\vb{r}\up} \rangle \hat{f}^{\dg}_{\vb{r}\down}\hat{f}_{\vb{r}\down}
    + \hat{f}^{\dg}_{\vb{r}\up}\hat{f}_{\vb{r}\up} \langle \hat{f}^{\dg}_{\vb{r}\down}\hat{f}_{\vb{r}\down} \rangle
    - \langle \hat{f}^{\dg}_{\vb{r}\up}\hat{f}_{\vb{r}\up} \rangle \langle \hat{f}^{\dg}_{\vb{r}\down}\hat{f}_{\vb{r}\down} \rangle \\
    &\quad
    - \langle \hat{f}^{\dg}_{\vb{r}\up}\hat{f}_{\vb{r}\down} \rangle \hat{f}^{\dg}_{\vb{r}\down}\hat{f}_{\vb{r}\up}
    - \hat{f}^{\dg}_{\vb{r}\up}\hat{f}_{\vb{r}\down} \langle \hat{f}^{\dg}_{\vb{r}\down}\hat{f}_{\vb{r}\up} \rangle
    + \langle \hat{f}^{\dg}_{\vb{r}\up}\hat{f}_{\vb{r}\down} \rangle \langle \hat{f}^{\dg}_{\vb{r}\down}\hat{f}_{\vb{r}\up} \rangle.
\end{aligned}
\end{equation}
Using a three-sublattice magnetic unit cell, the mean-field Hamiltonian can be written in momentum space as
\begin{equation}
\begin{aligned}
\label{eq:Hamiltonian_MF}
\hat{\mathcal{H}}_{\mathrm{PAM,MF}} 
&= \sum_{\mathbf{k}\in\rm{FBZ}} \left\{ \hat{\mathbf{c}}^\dagger_\mathbf{k}\hat{H}^{cc}_\mathrm{MF}(\mathbf{k})\hat{\mathbf{c}}_\mathbf{k} + \hat{\mathbf{f}}^\dagger_\mathbf{k}\hat{H}^{ff}_\mathrm{MF}(\mathbf{k})\hat{\mathbf{f}}_\mathbf{k} \right. \\
&\quad\quad\quad \left. + \left( \hat{\mathbf{c}}^\dagger_\mathbf{k}\hat{H}^{cf}_\mathrm{MF}(\mathbf{k})\hat{\mathbf{f}}_\mathbf{k} + \mathrm{h.c.} \right) \right\} + \mathrm{Const.}
\end{aligned}
\end{equation}
where,
\begin{equation}
\begin{aligned}
\hat{\mathbf{c}}_\mathbf{k} &= \left( \hat{c}_{\mathbf{k}1\up},\hat{c}_{\mathbf{k}1\dn},\hat{c}_{\mathbf{k}2\up},\hat{c}_{\mathbf{k}2\dn},\hat{c}_{\mathbf{k}3\up},\hat{c}_{\mathbf{k}3\dn}, \right)^\top, \\
\hat{\mathbf{f}}_\mathbf{k} &= \left( \hat{f}_{\mathbf{k}1\up},\hat{f}_{\mathbf{k}1\dn},\hat{f}_{\mathbf{k}2\up},\hat{f}_{\mathbf{k}2\dn},\hat{f}_{\mathbf{k}3\up},\hat{f}_{\mathbf{k}3\dn} \right)^\top,
\end{aligned}
\end{equation}
and
\begin{equation}
\begin{aligned}
    \hat{H}^{cc}_\mathrm{MF}(\mathbf{k}) &= -t \left(
    \begin{array}{ccc}
        0 & \gamma_{12}(\mathbf{k}) & \gamma^*_{31}(\mathbf{k}) \\
        \gamma^*_{12}(\mathbf{k}) & 0 & \gamma_{23}(\mathbf{k}) \\
        \gamma_{31}(\mathbf{k}) & \gamma^*_{23}(\mathbf{k}) & 0
    \end{array} \right) \otimes \sigma^0, \\
    \hat{H}^{ff}_\mathrm{MF}(\mathbf{k}) &= E_f \hat{\mathsf{1}}_{6\times6} + U\mathrm{diag}\left( \hat{\Sigma}_1,\hat{\Sigma}_2,\hat{\Sigma}_3 \right), \\
    \hat{H}^{cf}_\mathrm{MF}(\mathbf{k}) &= -V \hat{\mathsf{1}}_{6\times6}.
\end{aligned}
\end{equation}
The hopping matrix elements are
\begin{equation}
\begin{aligned}
\gamma_{12}(\mathbf{k}) &= 1+e^{-\mathrm{i}\mathbf{k}\cdot\mathbf{a}_3}+e^{+\mathrm{i}\mathbf{k}\cdot\mathbf{a}_2}, \\
\gamma_{23}(\mathbf{k}) &= 1+e^{-\mathrm{i}\mathbf{k}\cdot\mathbf{a}_1}+e^{+\mathrm{i}\mathbf{k}\cdot\mathbf{a}_3}, \\
\gamma_{31}(\mathbf{k}) &= 1+e^{-\mathrm{i}\mathbf{k}\cdot\mathbf{a}_2}+e^{+\mathrm{i}\mathbf{k}\cdot\mathbf{a}_1},
\end{aligned}
\end{equation}
while the mean-field potentials are given by
\begin{align}
    \hat{\Sigma}_\ell &= \left( \begin{array}{cc}
          \langle \hat{f}^\dagger_{\mathbf{k}\ell \dn}\hat{f}_{\mathbf{k}\ell \dn} \rangle &
        - \langle \hat{f}^\dagger_{\mathbf{k}\ell \dn}\hat{f}_{\mathbf{k}\ell \up} \rangle \\
        - \langle \hat{f}^\dagger_{\mathbf{k}\ell \up}\hat{f}_{\mathbf{k}\ell \dn} \rangle &
          \langle \hat{f}^\dagger_{\mathbf{k}\ell \up}\hat{f}_{\mathbf{k}\ell \up} \rangle
    \end{array} \right),
\end{align}
where $\ell$ ($=1,2,3$) labels the sublattices. The lattice vectors connecting neighboring magnetic unit cells are $\mathbf{a}_1=(0,\sqrt{3})$, $\mathbf{a}_2=(-3/2,-\sqrt{3}/2)$, and $\mathbf{a}_3=(3/2,-\sqrt{3}/2)$.

Diagonalizing the mean-field Hamiltonian in Eq.~(\ref{eq:Hamiltonian_MF}) for each momentum $\mathbf{k}$ yields the single-particle eigenvalues $\{\varepsilon_{\mathbf{k}\nu}\}$ and eigenstates $\{\ket{\mathbf{k}\nu}\}$, where $\nu$ ($=1, \cdots, 12$) labels the energy bands. The expectation values $\langle \hat{f}^{\dg}_{\vb{r}\sigma}\hat{f}_{\vb{r}\sigma'} \rangle$ are then updated self-consistently until the change in every mean field becomes smaller than $10^{-8}$.

Mean-field phase diagram of the periodic Anderson model in the plane of the onsite Coulomb repulsion $U$ and the onsite hybridization $V$ was obtained by Hayami \textit{et~al.}~\cite{Hayami2011}. Figure~\ref{Fig02}(a) shows the phase diagram at half filling when $E_f=-1$. For small hybridization $V$, the RKKY interaction, which scales as $V^2/U$, stabilizes the noncollinear $120^\circ$ antiferromagnetic state. In contrast, sufficiently large hybridization leads to complete Kondo screening at every site, resulting in a Kondo-insulating ground state. Between these two phases, a PD phase emerges owing to the competition between the RKKY interaction and the Kondo screening. The magnetic unit cell of the PD phase consists of three sublattices, in which one sublattice is nonmagnetic because of complete Kondo screening, whereas the remaining two exhibit N\'{e}el antiferromagnetic order [see Fig.~\ref{Fig01}(a)].

\subsection{Random-Phase Approximation}
Quantum fluctuations are completely neglected within the Hartree-Fock approximation, but they are expected to be important in the vicinity of phase boundaries. They also play an essential role in collective spin excitations through the formation of particle-hole bound states. To incorporate these effects beyond the mean-field approximation, we employ the random phase approximation (RPA) built on top of the self-consistent Hartree-Fock (SCHF) saddle-point solution as the reference state.

The generalized bare susceptibility is constructed from the Hartree-Fock eigenvalues $\{\varepsilon_{\mathbf{k}\nu}\}$ as,
\begin{equation}
\begin{aligned}
    \left[\hat{\chi}^0(\mathbf{q},\mathrm{i}\omega_n)\right]_{\sigma_1\sigma_2\sigma_3\sigma_4}^{abcd}
    &= -\frac{1}{N} \sum_{\mathbf{k}\in\rm{FBZ}} \sum_{\nu_1\nu_2} \frac{f(\varepsilon_{\mathbf{k}\nu_1}-\mu)-f(\varepsilon_{\mathbf{k}+\mathbf{q}\nu_2}-\mu)}{\mathrm{i}\omega_n+\varepsilon_{\mathbf{k}\nu_1}-\varepsilon_{\mathbf{k}+\mathbf{q}\nu_2}} \\
    &\quad\quad \times 
    u^{a\sigma_1}_{\mathbf{k}\nu_1} u^{b\sigma_2*}_{\mathbf{k}\nu_1} u^{c\sigma_3}_{\mathbf{k}+\mathbf{q}\nu_2} u^{d\sigma_4*}_{\mathbf{k}+\mathbf{q}\nu_2},
\end{aligned}
\end{equation}
where $u^{a\sigma}_{\mathbf{k}\nu}$ = $\langle a\sigma|\mathbf{k}\nu\rangle$, and $f(\varepsilon)=\left[ 1+\exp(\beta\varepsilon) \right]^{-1}$ is the Fermi-Dirac distribution function, $\mu$ is the chemical potential, and $N$ is the number of magnetic unit cells. The orbital indices $a$, $b$, $c$ and $d$ ($=c_1, c_2, c_3, f_1, f_2, f_3$) specify the conduction and localized $f$ orbitals within a magnetic unit cell.

The RPA susceptibility is obtained by resumming the infinite series of bubble diagrams,
\begin{equation}
\hat{\chi}^\mathrm{RPA}(\mathbf{q},\mathrm{i}\omega_n) = \hat{\chi}^0(\mathbf{q},\mathrm{i}\omega_n) \left[ \mathsf{1}+\hat{U}\hat{\chi}^0(\mathbf{q},\mathrm{i}\omega_n) \right]^{-1},
\end{equation}
The interaction-vertex tensor $\hat{U}$ has finite components 
\begin{align}
\left[U\right]^{abcd}_{\up\dn\dn\up}
=\left[U\right]^{abcd}_{\dn\up\up\dn}
=-\left[U\right]^{abcd}_{\up\up\dn\dn}
=-\left[U\right]^{abcd}_{\dn\dn\up\up}
=U,
\end{align}
when $a=b=c=d=f_\ell$ ($\ell=1,2,3$), while other components are zero. Our primary interest is the dynamical spin correlation functions of the localized spins,
\begin{equation}
\label{eq:defChi}
\hat{S}^{ff}_{\gamma\gamma'}(\mathbf{r},\mathbf{r}';\mathrm{i}\omega_n) = \int_{0}^\beta d\tau \Big\langle \mathcal{T}\left[\hat{S}_{\vb{r}}^\gamma(\tau)\hat{S}_{\vb{r}'}^{\gamma'}(0)\right]\Big\rangle \exp\left[\mathrm{i}\omega_n\tau \right]
\end{equation}
where $\hat{S}^a_\mathbf{r}=(1/2)\hat{f}^\dagger_{\mathbf{r}\sigma}\sigma^\gamma_{\sigma\sigma'}\hat{f}_{\mathbf{r}\sigma'}$ $(\gamma=x,y,z)$ is the localized spin operator. Here $\sigma^\gamma$ denotes the Pauli matrices, $\tau$ is the imaginary time, $\mathcal T$ is the imaginary-time ordering operator, $\beta=1/T$ is the inverse temperature, and $\omega_n$ is the bosonic Matsubara frequency.

Performing Wick decomposition followed by Fourier transformation, we obtain the transverse (in-plane) and longitudinal (out-of-plane) dynamical spin structure factors as
\begin{equation}
\begin{aligned}
& S^{ff}_\perp(\mathbf{q},\mathrm{i}\omega_n) = -\frac{1}{2\pi}\mathrm{Im} \sum_{\ell,m=f_1, f_2, f_3} \sum_\sigma
    \left[ \hat{\chi}^\mathrm{RPA}(\mathbf{q},\mathrm{i}\omega_n) \right]_{\sigma\sigma\bar{\sigma}\bar{\sigma}}^{\ell m m \ell}
    e^{-\mathrm{i}\mathbf{q}\cdot(\bm{\delta}_m-\bm{\delta}_\ell)}, \\
&\quad S^{ff}_\parallel(\mathbf{q},\mathrm{i}\omega_n) \\
& = -\frac{1}{4\pi}\mathrm{Im} \sum_{\ell,m=f_1, f_2, f_3} \sum_{\sigma,\sigma'}
    (-1)^{1-\delta_{\sigma\sigma'}}
    \left[ \hat{\chi}^\mathrm{RPA}(\mathbf{q},\mathrm{i}\omega_n) \right]_{\sigma'\sigma\sigma\sigma'}^{\ell m m \ell}
    e^{-\mathrm{i}\mathbf{q}\cdot(\bm{\delta}_m-\bm{\delta}_\ell)}.
\end{aligned}
\end{equation}
where $\bm{\delta}_1=(0,0)$, $\bm{\delta}_2=(0,1)$, and $\bm{\delta}_3=(1/2,\sqrt{3}/2)$ are the internal coordinates of the three sublattices within the magnetic unit cell. Finally, we perform the analytic continuation $\mathrm{i}\omega_n\rightarrow\omega+\mathrm{i}0^+$, which yields the experimentally observable dynamical spin structure factors as functions of the real frequency $\omega$.

The SCHF+RPA framework was originally introduced to describe spin excitations in cuprate superconductors~\cite{Schrieffer1988,Schrieffer1989} and was subsequently extended to iron-based superconductors~\cite{Knolle2011_itinerantSDW,Knolle2011_multiorbital,Knolle2011_magneticresonance}, where it successfully reproduces the gapless Goldstone modes associated with spontaneous symmetry breaking. The same framework has also been generalized to Abrikosov-fermion mean-field theories, including the staggered-flux phase on the square lattice~\cite{CMHo2001} and the Dirac spin-liquid ansatz for the triangular-lattice $J_1$-$J_2$ Heisenberg model~\cite{Willsher2025arXiv}. It has also been applied to the Haldane-Hubbard model, where it predicts a topological exciton-condensation-driven transition from a paramagnetic to a magnetically ordered phase~\cite{Eto2026arXiv}. Building on these developments, the present work extends the SCHF+RPA framework to partially disordered Kondo-lattice systems containing both magnetically ordered and Kondo-screened sites.

\section{Results}

\begin{figure*}[hbt]
    \centering
    \includegraphics[width=0.95\textwidth]{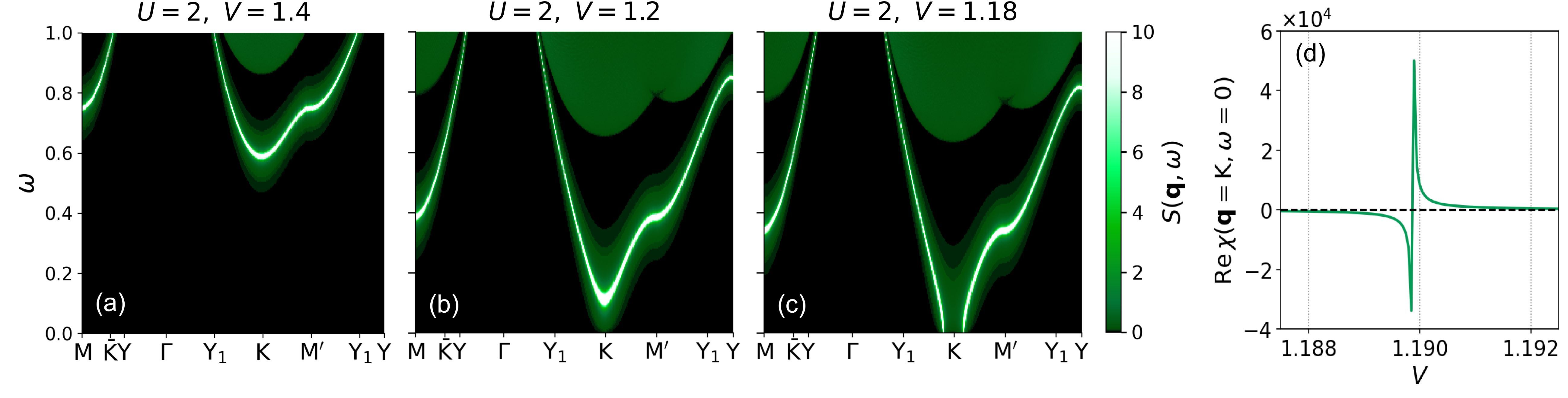}   
    \caption{
        (a)-(c) Transverse dynamical spin-structure factor $S^{ff}_\perp(\mathbf{q},\omega)$ assuming the KI reference state at $U=2$ for all panels, (a) $V=1.4$, (b) $V=1.2$, and (c) $V=1.18$.
        (d) $V$-dependence of the real part of the static RPA susceptibility at K point and at $U=2$.
    }
    \label{Fig03}
\end{figure*}

\begin{figure*}[hbt]
    \centering
    \includegraphics[width=0.95\textwidth]{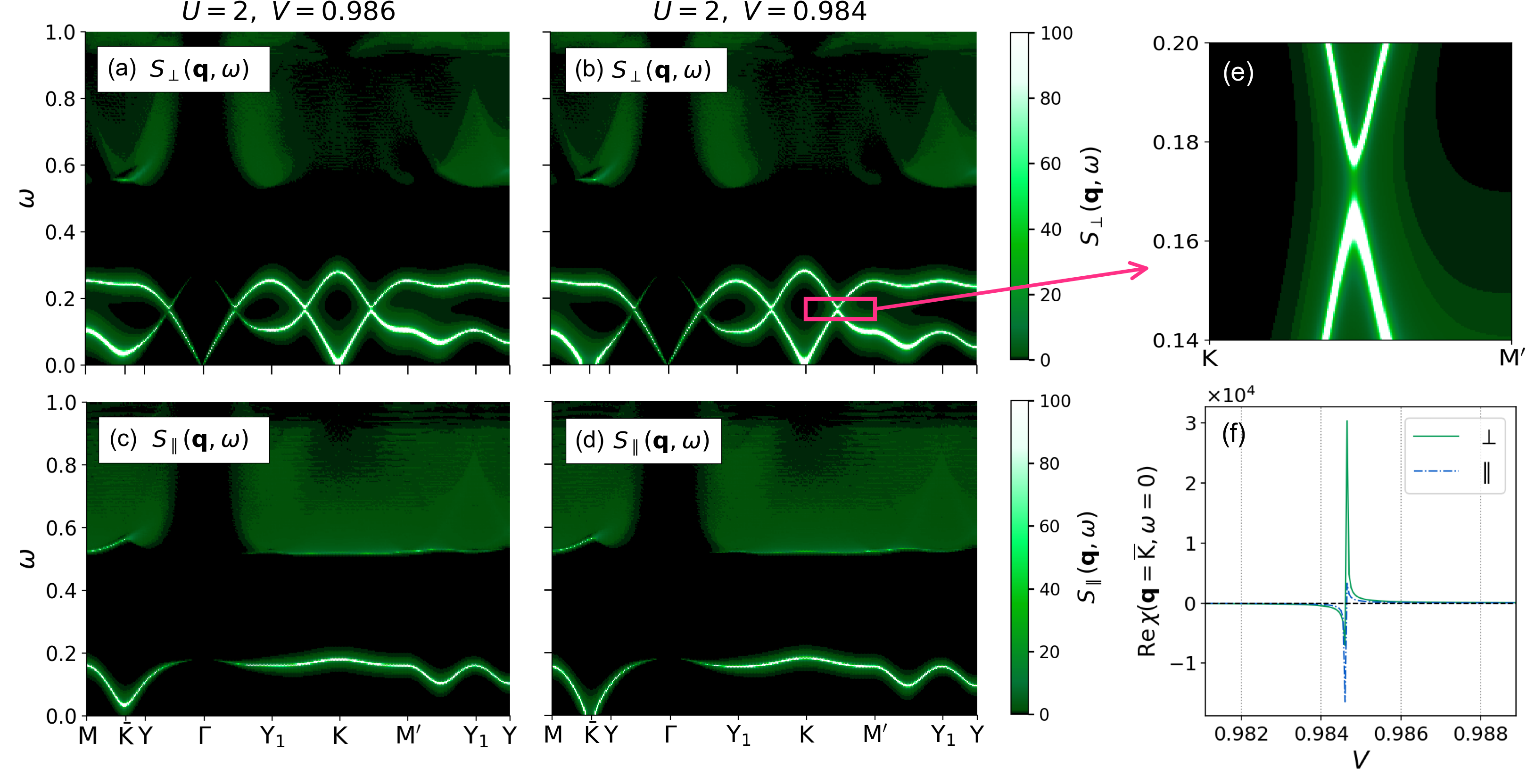}   
    \caption{
        (a)-(d) (a),(b) Transverse and (c),(d) longitudinal dynamical spin-structure factor $S^{ff}_{\perp/\parallel}(\mathbf{q},\omega)$ assuming the three-sublattice PD mean-field reference state at $U=2$ for all panels, (a),(c) $V=0.986$ and (b),(d) $V=0.984$.
        (e) Enlarged view of the transverse dynamical spin-structure factor at $(U,V)=(2,0.984)$. The plotted momentum-frequency region corresponds to the magenta box in (b).
        (f) $V$-dependence of the real part of the static RPA susceplibility at $\bar{\mathrm{K}}$ point and at $U=2$.
    }
    \label{Fig04}
\end{figure*}

\subsection{Triplons as particle-hole bound states}

Figure~\ref{Fig03}(a) shows the transverse dynamical spin-structure factor deep inside the KI phase $(U,V)=(2,1.4)$ along the high-symmetry momentum path presented in Fig.~\ref{Fig01}(b). Note that, since the KI phase preserves the spin-rotation symmetry of Eq.~(\ref{eq:PAM_Hamiltonian}), the structure factor satisfies the relation $S^{ff}_\perp(\mathbf{q},\omega) = 2S^{ff}_\parallel(\mathbf{q},\omega)$.
A sharp and dispersive low-energy mode is found below the particle-hole continuum, which can be understood as a particle-hole bound state corresponding to a $\Delta S=1$ spin-flip excitation from the underlying singlet ground state.
Such a single spin-flip excitation out of a singlet ground state is usually called a triplet excitation, or a triplon in a quasiparticle picture. This triplon mode is gapped over the entire Brillouin zone, as indeed observed experimentally in the Kondo insulator SmB${}_6$~\cite{Fuhrman2015,Fuhrman2018}. 
These results show that the low-energy dynamics of the KI phase is largely governed by triplons.
It is worth pointing out that the vanishing spectral weight in the long-wavelength limit $\mathbf{q}\rightarrow\Gamma$ highlights antiferromagnetic correlations among neighboring localized spins.

\subsection{Phase transitions driven by triplon Bose-Einstein condensation}

We next investigate how the triplon excitation in the KI phase evolves toward the transition into the PD phase.
Figures~\ref{Fig03}(b) and~\ref{Fig03}(c) show the transverse dynamical spin-structure factor calculated within the RPA by taking the KI solution as the reference state.
Although the parameter set of Fig.~\ref{Fig03}(c) already lies inside the PD region of the mean-field phase diagram, the RPA calculation is still performed around the metastable nonmagnetic KI solution.
Therefore, the spectrum shown in Fig.~\ref{Fig03}(c) describes the instability of the KI state rather than the collective excitations of the PD phase itself.

As the hybridization strength $V$ is reduced, the triplon gap at the K point continuously decreases and eventually closes.
The gap closing indicates that the KI state becomes unstable against the spontaneous condensation of triplons carrying momentum $\mathbf{q}=\mathrm{K}$.
The ordering wave vector of the condensed triplons coincides with that of the three-sublattice PD order, demonstrating that the KI-PD phase transition is driven by Bose-Einstein condensation of triplons.
This conclusion is fully consistent with the divergence of the static RPA susceptibility, $\mathrm{Re}\,\chi^{\mathrm{RPA}}(\mathbf{q}=\mathrm{K},0)$, presented in Fig.~\ref{Fig03}(d), which provides an equivalent criterion for the onset of the instability.
In the parameter region beyond the critical point, the triplon mode moves into the negative frequency regime.
This unphysical behavior indicates the breakdown of the KI reference state, implying that the true ground state is no longer the KI phase.
It is worth noting that, as seen in Fig.~\ref{Fig02}(b), the phase boundary determined from the triplon condensation agrees almost perfectly with that obtained from the SCHF calculation.
This remarkable agreement provides another strong piece of evidence that the KI-PD phase transition is indeed driven by the Bose-Einstein condensation of triplons.

Having identified the instability of the KI state, we now switch to the three-sublattice PD mean-field solution as the reference state and investigate its collective excitations within the RPA. Typical examples of $S^{ff}_\perp(\mathbf{q},\omega)$ and $S^{ff}_\parallel(\mathbf{q},\omega)$ in the PD phase are presented in Figs.~\ref{Fig04}(a) and~\ref{Fig04}(c), respectively. 
In the PD phase, spin-rotation symmetry is spontaneously broken from SO(3) to SO(2), with the residual SO(2) symmetry corresponding to rotations about the N\'{e}el vector.
Consequently, the transverse and longitudinal channels are no longer equivalent.
Two degenerate, linearly dispersive gapless Goldstone modes appear in the transverse channel, whereas no corresponding gapless mode is present in the longitudinal channel.
These Goldstone modes can be identified as magnons whose spectral weight is concentrated predominantly on the magnetically ordered sublattices.
In addition, the transverse spectrum contains a gapped triplon-like mode throughout the Brillouin zone, with spectral weight concentrated mainly on the Kondo-screened sublattice.
A related gapped mode also appears in the longitudinal channel, although its dispersion differs from that of the transverse mode, reflecting the broken spin-rotation symmetry of the PD state.

As the hybridization strength $V$ is reduced, Figs.~\ref{Fig04}(b) and~\ref{Fig04}(d) show that the gapped triplon-like modes at the $\bar{\mathrm{K}}$ point, which is located at a corner of the three-sublattice magnetic Brillouin zone, gradually soften and eventually condense.
Although the dispersions of the triplon-like modes in the transverse and longitudinal channels are no longer identical because of the spontaneous breaking of the SO(3) spin-rotation symmetry, their excitation gaps at the $\bar{\mathrm{K}}$ point vanish simultaneously at the same critical value of $(U,V)$. Reflecting this coincidence, both the transverse and longitudinal channels of the static RPA susceptibility presented in Fig.~\ref{Fig04}(f) diverge simultaneously within negligible numerical error. This simultaneous gap closing indicates that the condensed excitation branch retains the underlying partial-singlet character of the PD state, further supporting its triplon-like nature.
Thus, the onset of the instability is independent of the spin channel despite the inequivalent excitation spectra away from the critical point.
The simultaneous condensation of the triplon-like modes at $\bar{\mathrm{K}}$ signals an instability of the three-sublattice PD state toward a nine-sublattice magnetic order.
This instability is absent from previous mean-field studies restricted to the three-sublattice unit cell~\cite{Hayami2011}, but becomes apparent from our collective-mode analysis within the SCHF+RPA.
To further examine the nature of this instability, in Fig.~\ref{Fig02}(c), we compare the mean-field free energies of the PD state and a candidate nine-sublattice noncoplanar magnetic state illustrated in Fig.~\ref{Fig02}(b).
This state can be viewed as a three-sublattice PD state supplemented by a weak \(120^\circ\) magnetic order developing on the originally Kondo-screened sublattice, with the additional moments lying in the plane perpendicular to the Néel vector of the PD state. This structure is fully consistent with the simultaneous condensation of the transverse and longitudinal triplon-like modes discussed above, indicating that the instability primarily originates from the Kondo-screened sublattice and induces magnetic ordering on the sites that remain screened in the PD phase.
Remarkably, the phase boundary determined from the crossing of their SCHF free energies coincides with the $\bar{\mathrm{K}}$-point triplon-condensation line obtained from the SCHF+RPA [see Fig.~\ref{Fig02}(c)]. This agreement provides independent support for identifying the soft mode with an instability toward the nine-sublattice state. Moreover, on the small-$V$ side of this boundary, the nine-sublattice state has a lower mean-field free energy than the conventional three-sublattice $120^\circ$ magnetic state. Although its detailed spin configuration remains to be fully characterized, the combined RPA instability and mean-field energetics lead us to propose this nine-sublattice magnetic state as a new candidate ground state of the model in the small-$V$ regime.

Finally, we discuss the hybridization between the magnon-like and triplon-like modes in the PD phase.
Since both magnons and triplons carry the same spin quantum number, $\Delta S=1$, their mutual hybridization is symmetry-allowed, at least at generic crystal momenta where the little group consists only of the identity operation.
Indeed, Fig.~\ref{Fig04}(e) shows the resulting hybridization gap clearly.
We stress that this hybridization is an observable feature unique to the PD state in which magnon and triplon excitations coexist.

\section{Discussion}

\subsection{Limitation of SCHF+RPA}

A limitation of the present SCHF+RPA approach is that quantum fluctuations are not incorporated at the level of the SCHF reference state itself, since the RPA is constructed on top of a static Hartree-Fock saddle point. Consequently, quantities such as ordered moments, excitation energies, and phase boundaries may not be quantitatively accurate when quantum fluctuations strongly renormalize the underlying ground state. On the other hand, an important advantage of this framework is that the dynamical spin correlations are determined directly from the microscopic Hamiltonian and the self-consistent mean-field solution, without introducing additional phenomenological fitting parameters. A more quantitative description may be obtained by replacing the Hartree-Fock reference state with a more correlated one, for example within a slave-boson mean-field framework~\cite{Peres2002slave,Issing2026arXiv}.

\subsection{Comparison with nonperturbative simulations}

The stability of the PD phase obtained within the present SCHF+RPA framework can be further discussed by comparison with previous nonperturbative studies. Previous DMFT calculations for the triangular-lattice periodic Anderson model reported that the PD phase is not stabilized at half filling, but instead emerges upon hole doping between magnetically ordered and Kondo-screened phases~\cite{Aulbach2015}. This is in contrast to the present SCHF calculation, where the three-sublattice PD state is stabilized already at half filling.
This discrepancy likely reflects the complementary aspects of electronic correlations captured by the two approaches. DMFT treats local dynamical correlations and the frequency dependence of the electronic self-energy nonperturbatively, and therefore provides a more accurate description of the local quantum fluctuations associated with Kondo screening. In contrast, the present SCHF+RPA approach is constructed on top of a static Hartree-Fock reference state and does not incorporate such dynamical self-energy effects at the saddle-point level. On the other hand, the present framework retains full momentum resolution in the collective response and can directly resolve spatially modulated magnetic instabilities. The \(\bar{\mathrm{K}}\)-point instability toward the nine-sublattice magnetic state found in this work provides a clear example of such an instability, which cannot be captured within a calculation restricted to the original three-sublattice periodicity.
The different conclusions concerning the PD phase at half filling may therefore originate from the different treatment of temporal and spatial correlations. The DMFT result suggests that local dynamical correlations can substantially modify the stability of the mean-field PD state, whereas the present results demonstrate that momentum-dependent collective fluctuations can generate additional ordering tendencies beyond the assumed mean-field unit cell. Since partial disorder intrinsically involves both local Kondo screening and spatially inhomogeneous magnetic ordering, a complete description of its stability ultimately requires both aspects to be treated on an equal footing.

A complementary perspective is provided by many-variable variational Monte Carlo (mVMC) calculations for the frustrated Kondo-lattice model, where a partially Kondo-screened ground state was found between magnetically ordered and fully Kondo-screened regimes~\cite{Motome2010}. Although the Kondo-lattice model differs microscopically from the periodic Anderson model studied here, the mVMC results demonstrate that partial Kondo screening can remain stable even when quantum fluctuations are treated beyond static mean-field theory. This indicates that the emergence of a PD state is not intrinsically tied to the Hartree-Fock approximation. At the same time, because the two studies are based on different microscopic models, a direct quantitative comparison of their phase boundaries is not possible.

Taken together, these nonperturbative studies indicate that the stability of the PD state is sensitive to the interplay between local screening and spatial magnetic correlations. The present SCHF+RPA approach complements these methods by providing direct access to the momentum- and frequency-resolved collective excitations without introducing any phenomenological fitting parameters. In particular, it enables us to identify not only the triplon condensation responsible for the formation of the PD state, but also its subsequent instability toward a larger-unit-cell noncoplanar magnetic order. A theoretical framework combining such spatial resolution with a fully dynamical treatment of the electronic self-energy is an important direction for future work.

\section{Conclusion}

In summary, we have investigated the collective spin excitations and stability of the partially disordered phase in the half-filled periodic Anderson model on a triangular lattice using the self-consistent Hartree-Fock plus random-phase approximation. We showed that the transition from the Kondo-insulating phase to the PD phase is driven by the condensation of a triplon excitation associated with the Kondo singlet. In the PD phase, gapless magnon modes originating predominantly from the magnetically ordered sublattices coexist and hybridize with gapped triplon-like modes associated mainly with the Kondo-screened sublattice, giving rise to a characteristic level repulsion in the excitation spectrum.
We further identified a second triplon-like condensation at the \(\bar{\mathrm{K}}\) point, signaling an instability of the three-sublattice PD state toward a nine-sublattice magnetic state. The corresponding nine-sublattice noncoplanar mean-field solution can be viewed as a PD state in which a weak \(120^\circ\) magnetic order develops on the originally Kondo-screened sublattice. The SCHF phase boundary between the PD and the nine-sublattice state perfectly coincides with the \(\bar{\mathrm{K}}\)-point condensation line obtained from the RPA, and the mean-field free energy of the nine-sublattice state becomes lower than that of the conventional three-sublattice \(120^\circ\) state on the small-\(V\) side. These results identify the nine-sublattice state as a new candidate ground state in the small-\(V\) regime.

Our results demonstrate that the low-energy collective excitations provide direct insight into both the formation and the instability of partially disordered states. More broadly, they establish a microscopic picture in which Kondo-singlet excitations and magnetic Goldstone modes coexist, hybridize, and ultimately drive successive magnetic instabilities in frustrated heavy-fermion systems.

An interesting direction for future work is to extend the present analysis to real materials such as UNi$_4$B, where both toroidal and triple-$Q$ triforce orders have been discussed as candidate ground states~\cite{Hayami2015,Ishitobi2023}. It would be interesting to clarify how magnon and triplon-like excitations evolve in such unconventional magnetic and orbital orders.

\section*{Acknowledgment}
M.M. acknowledges support by the CREST, the Japan Science and Technology Agency (No. JPMJCR20T1) and support by the Grants-in-Aid for Scientific Research from JSPS KAKENHI (No. JP25H00611, and No. JP24H02231) and the Waseda University Grant for Special Research Projects (No. 2026C147).
R.E. was supported by Grant-in-Aid for JSPS Fellows (Grant No. 23KJ2047).
R.E. further acknowledges financial support from JSPS through Overseas Postdoctoral Fellowship.

\bibliography{refs}
	
\end{document}